&pdflatex

\documentclass[10pt,twocolumn,prd,aps,amssymb,amsmath,tightenlines,showpacs]{revtex4}
\usepackage{graphicx}

\newcommand{\veps}{\varepsilon}

\newcommand{\cP}{\ensuremath{\mathcal{P}}}
\newcommand{\cT}{\ensuremath{\mathcal{T}}}
\newcommand{\half}{\mbox{$\textstyle{\frac{1}{2}}$}}

\begin{document}

\title{Observation of $\cP\cT$ phase transition in a simple mechanical system}

\author{Carl~M.~Bender${}^1$\footnote{Permanent address: Department of
Physics, Washington University, St. Louis, MO 63130, USA.\\ email:
cmb@wustl.edu}, Bjorn K.~Berntson${}^2$\footnote{email:
bjorn.berntson11@imperial.ac.uk}, David Parker${}^1$\footnote{email:
david.j.parker@kcl.ac.uk}, and
E.~Samuel${}^1$\footnote{email: ernest.samuel@kcl.ac.uk}}

\affiliation{${}^1$Department of Physics, King's College London, Strand,
London WC2R 1LS, UK\\
${}^2$Blackett Laboratory, Imperial College, London SW7 2AZ, UK}

\date{\today}

\begin{abstract}
If a Hamiltonian is $\cP\cT$ symmetric, there are two possibilities: Either the
eigenvalues are entirely real, in which case the Hamiltonian is said to be in an
{\it unbroken-$\cP\cT$-symmetric phase}, or else the eigenvalues are partly real
and partly complex, in which case the Hamiltonian is said to be in a {\it
broken-$\cP\cT$-symmetric phase}. As one varies the parameters of the
Hamiltonian, one can pass through the phase transition that separates the
unbroken and broken phases. This transition has recently been observed in a
variety of laboratory experiments. This paper explains the phase transition in a
simple and intuitive fashion and then describes an extremely elementary
experiment in which the phase transition is easily observed.
\end{abstract}
\maketitle

\section{Introduction}
\label{s1}

Quantum-mechanical $\cP\cT$-symmetric Hamiltonians often exhibit two parametric
regions, a region of {\it unbroken} $\cP\cT$ symmetry in which the eigenvalues
are all real, and a region of {\it broken} $\cP\cT$ symmetry in which some of
the eigenvalues are real and the remaining eigenvalues are complex. (Here, $\cP$
represents space reflection, or parity, and $\cT$ represents time reversal.) A
phase transition occurs at the boundary between these two regions. This paper
describes an elementary experiment that demonstrates the $\cP\cT$ phase
transition in a simple $\cP\cT$-symmetric classical-mechanical system involving
two coupled driven pendula.

If a quantum-mechanical Hamiltonian is Dirac Hermitian (that is, if $H=H^\dag$,
where $H^\dag$ is the transpose and complex conjugate of $H$), then its energy
eigenvalues are always real. Thus, a Dirac Hermitian Hamiltonian can never
exhibit a phase transition where its eigenvalues go from being real to being
complex. However, in 1998 it was shown that the eigenvalues of a $\cP
\cT$-symmetric Hamiltonian can be entirely real, even if the Hamiltonian is not
Dirac Hermitian \cite{r1,r2,r3}. The class of $\cP\cT$-symmetric Hamiltonians
considered in Refs.~\cite{r1,r2,r3} has the form
\begin{equation}
H=p^2+x^2(ix)^\veps,
\label{e1}
\end{equation}
where $\veps$ is a real parameter. These Hamiltonians are not Dirac Hermitian
(except at $\veps=0$), but they are $\cP\cT$ symmetric because under space
reflection $x$ changes sign and under time reversal $i$ changes sign. The
Hamiltonians (\ref{e1}) exhibit a region of unbroken $\cP\cT$ symmetry ($\veps
\geq0$) and a region of broken symmetry ($-1<\veps<0$). Thus, at $\veps=1$ and
at $\veps=2$ we obtain the Hamiltonians 
\begin{equation}
H=p^2+ix^3
\label{e2}
\end{equation}
and
\begin{equation}
H=p^2-x^4,
\label{e3}
\end{equation}
and surprisingly, the eigenvalues of these Hamiltonians are all real, positive,
and discrete. However, at $\veps=-1/2$ we obtain the Hamiltonian $H=p^2+x^2(ix
)^{-1/2}$, which has an infinite number of complex eigenvalues and only a finite
number of real eigenvalues.

Recently, the $\cP\cT$ phase transition has been observed in laboratory
experiments on many different $\cP\cT$-symmetric systems
\cite{r4,r5,r6,r7,r8,r9,r10,r11,r12}. The coupled-electronic-oscillator
experiment by J.~Schindler {\it et al.} \cite{r11} is particularly elegant
because of its simplicity, and we were motivated by this experiment to construct
its mechanical analog by using a pair of coupled driven pendula. We have been
able to do this, and in this mechanical system the $\cP\cT$ phase transition is
easy to observe.

Mechanical oscillators (coupled pendula) have been used in other contexts to
model complex physical phenomena. For example, in particle physics, it was
proposed that transitions in ${\rm K}^0{\bar{\rm K}}^0$, ${\rm B}^0{\bar{\rm B}
}^0$, and ${\rm D}^0{\bar{\rm D}}^0$ systems could be visualized by using
coupled pendula \cite{r13}. These experiments have recently been performed
\cite{r14,r15}. In these experiments, systems of coupled, damped pendula were
used. The experiment reported in this paper differs from these experiments in
that while one of our pendula is damped, the other is {\it undamped} (driven).

This paper is organized as follows. Section \ref{s2} presents an intuitive
explanation of the $\cP\cT$ phase transition. The pendulum experiment is
described in Sec.~\ref{s3}. Section \ref{s4} gives concluding remarks and
proposes some follow-up experiments that we believe can be readily performed.

\section{Intuitive Explanation of the $\cP\cT$ Phase Transition}
\label{s2}

Originally, $\cP\cT$-symmetric quantum mechanics was studied at a highly
mathematical level. In Ref.~\cite{r1}, the techniques of complex variables,
asymptotics, differential equations, and perturbative theory were used to
understand the analytic continuation of eigenvalue problems and $\cP\cT$
symmetry. The rigorous proof that the eigenvalues of $H$ in (\ref{e1}) are real
and positive when $\veps\geq0$ involves the use of sophisticated mathematical
techniques (Baxter TQ relation, Bethe {\it ansatz}, functional determinants),
which are used to study integrable systems and conformal field theory
\cite{r16,r17}.

However, as experiments on $\cP\cT$-symmetric physical systems have been
performed and published, we have realized that the $\cP\cT$ phase transition can
be explained simply and intuitively without resorting to sophisticated
mathematics. We present this explanation below.

\begin{figure}[h!]
\begin{center}
\includegraphics[trim=7mm 7mm 1mm 1mm,clip=true,scale=0.36]{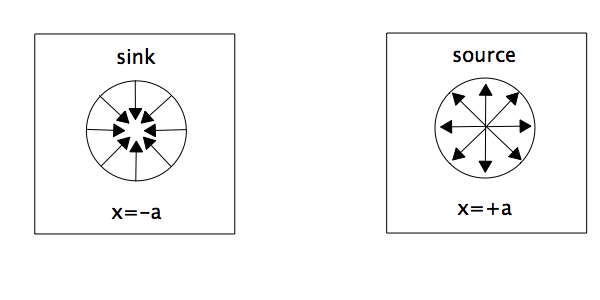}
\caption{A simple schematic $\cP\cT$-symmetric physical system: A box located at
$x=-a$ contains a sink (an antenna that absorbs) and a box located at $x=+a$
contains a source (an antenna that radiates at an equal rate). Under space
reflection $\cP$ the boxes interchange, and under time reversal $\cT$ the sink
becomes a source and the source becomes a sink. Thus, the system is $\cP\cT$
symmetric. If the boxes are isolated, the system cannot be in equilibrium
because the energy decays to zero in the left box and grows to infinity in the
right box. Thus, in this case the $\cP\cT$ symmetry of the system is broken.
The energies corresponding to these runaway modes are complex. However, if the
boxes are sufficiently strongly coupled, the system can equilibrate and the $\cP
\cT$ symmetry is unbroken. When the system is in equilibrium, its energy is
real.}
\label{F1}
\end{center}
\end{figure}

Imagine two identical closed and isolated boxes, one on the negative-$x$ axis at
$x=-a$, and the other on the positive-$x$ axis at $x=a$ (see Fig.~\ref{F1}).
Inside the left box is a sink (an antenna that absorbs energy) and inside the
right box is a source (an antenna that radiates energy at an equal rate). This
system is $\cP\cT$ symmetric because under parity $\cP$ the left box and the
right box interchange positions and under time reversal $\cT$ the sink becomes a
source and the source becomes a sink.

The Hamiltonian that describes the time evolution of the one-dimensional system
in the left box is the $1\times1$ matrix $H=[E_1]=\left[ae^{i\theta}\right]$,
where $a>0$ and $0<\theta<\pi$ so that ${\rm Im}\,E_1>0$. The solution to the
time-dependent Schr\"odinger equation for this system,
\begin{equation}
-i\frac{d}{dt}\phi(t)=H\phi(t),
\label{e4}
\end{equation}
is $\psi(t)=\psi(0)e^{iE_1t}$, which decays exponentially with time because
there is a sink in the left box.

Similarly, the Hamiltonian that describes the time evolution of the
one-dimensional system in the right box is the $1\times1$ matrix $H=[E_2]=\left[
ae^{-i\theta}\right]$, so that ${\rm Im}E_2<0$. The solution to the
time-dependent Schr\"odinger equation for the right system is $\psi(t)=\psi(0)
e^{iE_2t}$, which grows exponentially with time because there is a source in
the right box.

The left and right systems taken together are described by the $2\times2$
diagonal matrix Hamiltonian
\begin{equation}
H=\left[\begin{array}{cc} ae^{i\theta} & 0\\ 0 & ae^{-i\theta}\\ \end{array}
\right].
\label{e5}
\end{equation}
This Hamiltonian is not Dirac Hermitian and its eigenvalues $E_1$ and $E_2=
E_1^*$ are complex. However, it is $\cP\cT$ symmetric, where the parity operator
$\cP$ is the matrix
\begin{equation}
\cP=\left[\begin{array}{cc} 0&1\\ 1&0\\ \end{array}\right],
\label{e6}
\end{equation}
which interchanges the two systems, and the time reversal operator $\cT$ is
complex conjugation. The system described by $H$ in (\ref{e5}) is not in
equilibrium because the eigenfunction in the left box decays exponentially and
the eigenfunction in the right box grows exponentially. Correspondingly, $E_1$
and $E_2$, the eigenvalues of $H$, are complex. Thus, the system is in a
broken-$\cP\cT$-symmetric phase. 

The combined two-box system can achieve equilibrium if we couple the boxes
together sufficiently strongly. We couple the boxes symmetrically in order to
preserve the $\cP\cT$ symmetry of the entire system:
\begin{equation}
H=\left[\begin{array}{cc} ae^{i\theta} & g\\ g & ae^{-i\theta}\\ \end{array}
\right],
\label{e7}
\end{equation}
where the coupling constant $g$ is real. This coupling allows the excess energy
in the right box to leak into the left box.

The eigenvalues of $H$ in (\ref{e7}) become real if $g^2>a^2\sin^2\theta$. Thus,
if the coupling is strong enough, the excess energy in the left box can flow
into the right box fast enough for the system to remain in equilibrium. When
this happens, the energy eigenvalues are real and the system is in an
unbroken-$\cP\cT$-symmetric phase. The phase transition occurs when the coupling
constant $g$ exceeds the critical value given by
\begin{equation}
g_{\rm crit}^2=a^2\sin^2\theta.
\label{e8}
\end{equation}

Using this simple two-dimensional model, we can now understand heuristically why
the Hamiltonians in (\ref{e1}) have an unbroken $\cP\cT$ symmetry when $\veps>0$
and a broken $\cP\cT$ symmetry when $\veps<0$. Consider, for example, the
potential $ix^3$ for $H$ in (\ref{e2}); this potential has a positive-imaginary
part when $x>0$ and a negative-imaginary part when $x<0$. Thus, there is a
nonlocal source everywhere on the negative-$x$ axis and a corresponding nonlocal
sink everywhere on the positive-$x$ axis. Can such a system actually be in
equilibrium? One might think that it would be impossible for this system to
reach equilibrium because the source becomes infinitely strong as $x\to-\infty$
and the sink becomes infinitely strong as $x\to+\infty$. However, by means of a
simple classical argument, we can easily show that the system can indeed achieve
equilibrium: We determine {\it how long} it takes for particles at $x=-\infty$
to flow to $x=+\infty$. The classical time of flight $T$ is given by
\begin{equation}
T=\int dt=\int\frac{dx}{p}=\int_{x=-\infty}^\infty\frac{dx}{\sqrt{E-ix^3}},
\label{e9}
\end{equation}
where we have used the condition that the Hamiltonian represents the classical
energy: $E=p^2+ix^3$. Since the integral in (\ref{e9}) {\it converges}, the time
of flight from $x=-\infty$ to $x=+\infty$ is {\it finite}, and thus the system
can attain equilibrium. Clearly, if $\veps$ in (\ref{e1}) becomes negative, the
time of flight integral diverges and the system cannot be in equilibrium. This
is why the $\cP\cT$ symmetry of $H$ breaks and the eigenvalues become complex
when $\veps<0$.

This simple two-box source-and-sink model explains what was done in the
microwave cavity experiment in Ref.~\cite{r12}. In this experiment two coupled
microwave cavities, one containing a source of microwaves and the other
containing a sink of microwaves was used, and as the coupling was varied, the
$\cP\cT$ phase transition was observed.

The source-and-sink model also explains the two-channel optics experiments in
Refs.~\cite{r6} and \cite{r7}. In these experiments light travels down a
coupled pair of wave guides. One wave guide has loss and the other has gain. If
the coupling of the wave guides is sufficiently strong, then the system is in
equilibrium, and one observes {\it Rabi oscillations} (power oscillations) in
which the optical energy oscillates between the two wave guides. When the
coupling between the wave guides becomes too weak, the Rabi oscillations cease
and the system can no longer remain in equilibrium; the power grows
exponentially in one wave guide and decays exponentially in the other.

The experiment in Ref.~\cite{r11} is the electronic analog of the two-channel
optics experiments. The electronic experiment involves two
inductively-coupled LRC circuits, one with gain and the other with loss. Once
again, as the coupling between the two oscillators becomes weaker than a
critical value, the system enters the $\cP\cT$-broken phase and is no longer in
equilibrium.

\section{Description of the Experiment}
\label{s3}
The experiment discussed in this paper is a classical-mechanical analog of the
electronic experiment in Ref.~\cite{r11}. Instead of two LRC circuits, there are
two coupled pendula, one with energy loss and one with energy gain. We first
formulate some elementary equations to model a coupled a two-pendulum system and
then we describe the experiment itself.

\subsection{An overly simple mathematical model of a two-oscillator system}
\label{ss31}

\begin{widetext}

\begin{figure}[t!]
\begin{center}
\includegraphics[trim=0mm 10mm 0mm 5mm,clip=true,scale=0.315]{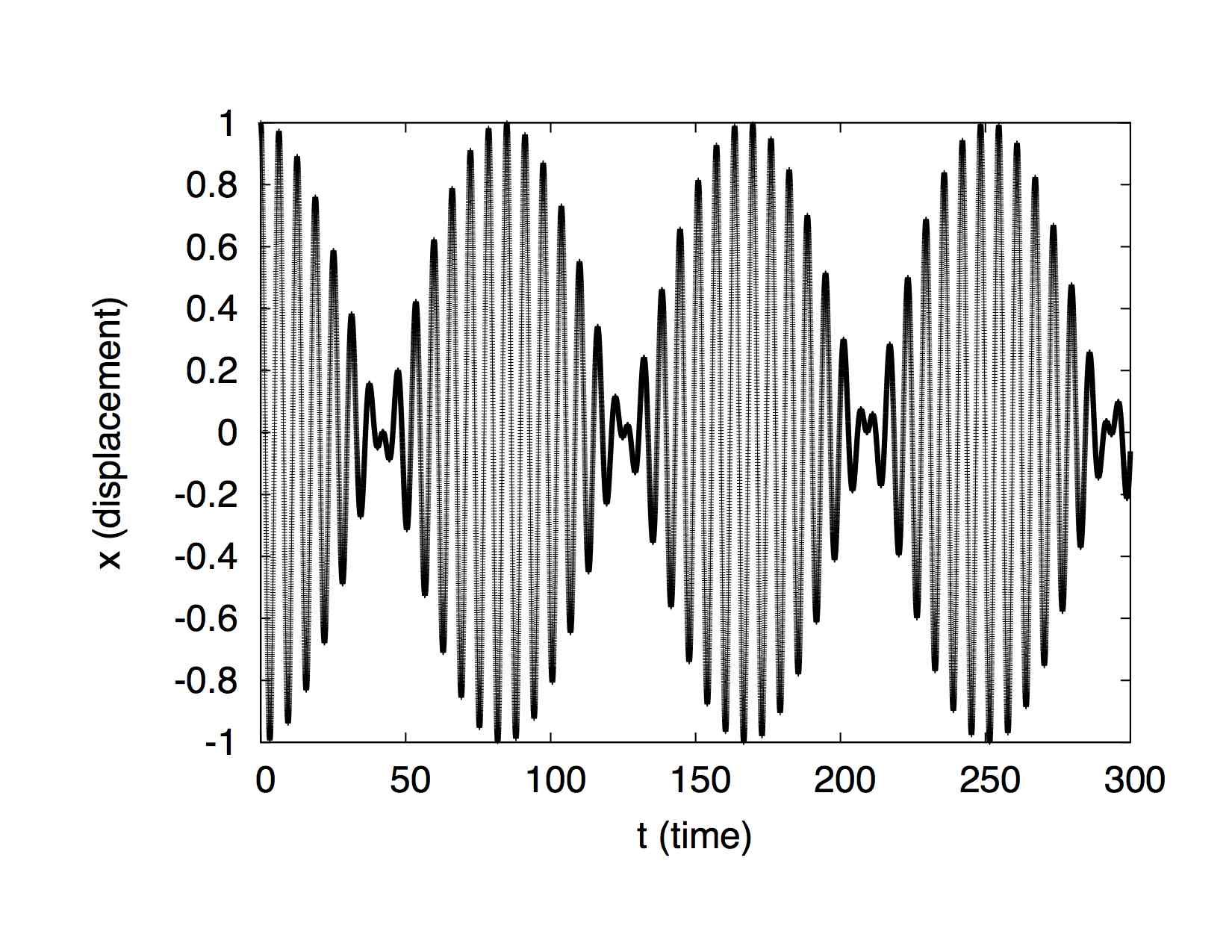}
\hspace{0.1cm}
\includegraphics[trim=0mm 10mm 0mm 5mm,clip=true,scale=0.315]{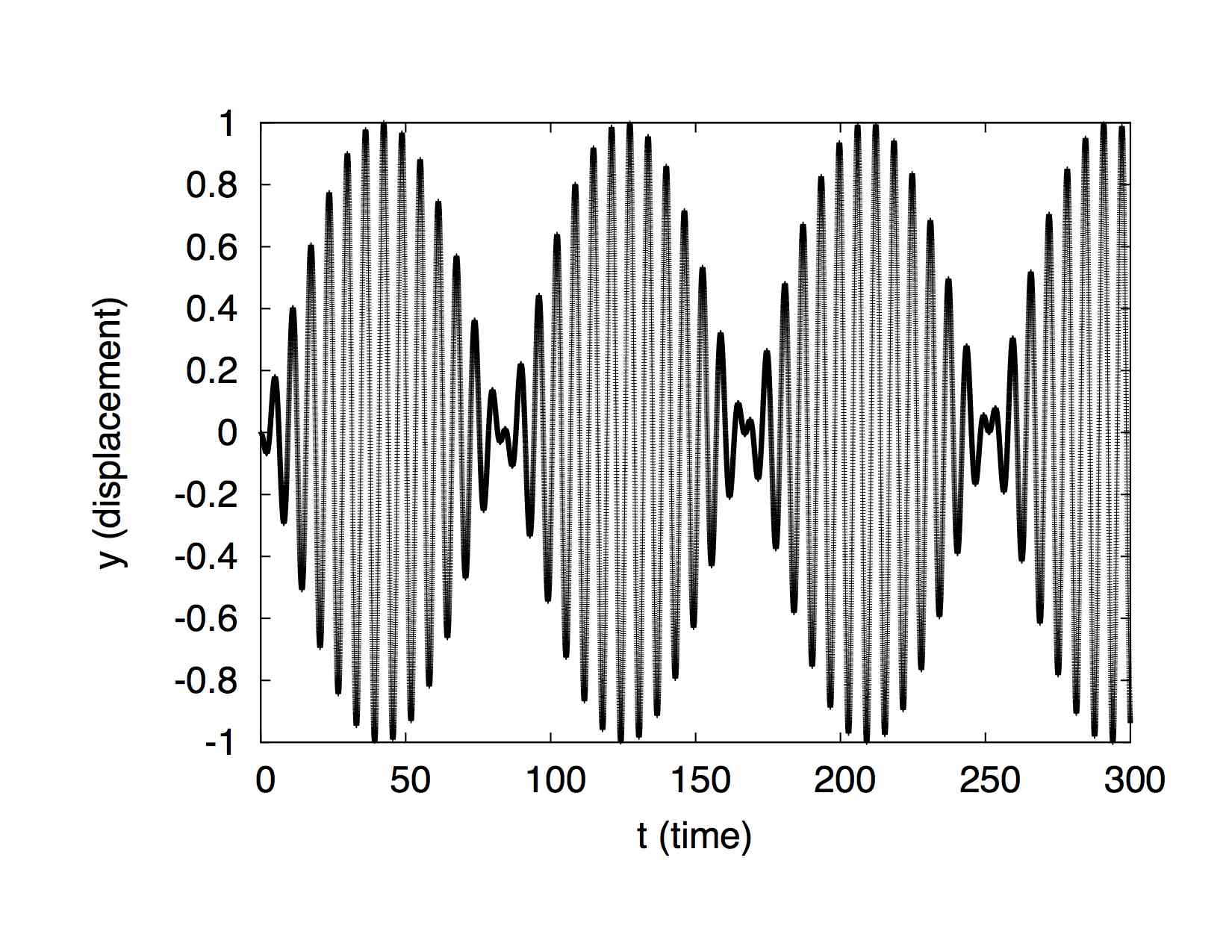}
\end{center}
\caption{Oscillatory motion of a pair of coupled pendula [see (\ref{e11})] with
no loss or gain. The $x$ displacement is shown in the left panel and the $y$
displacement is shown in the right panel. The coupling parameter is $\veps=
0.075$. Rabi power oscillations occur in which the maximum and minimum
displacements $x(t)$ and $y(t)$ are $90^\circ$ out of phase.}
\label{F2}
\end{figure}

\end{widetext}

The Hamiltonian
\begin{equation}
H=\half p^2+\half x^2+\half q^2+\half y^2+\veps xy
\label{e10}
\end{equation}
describes the small-amplitude frictionless motion of two coupled pendula, where
$x(t)$ and $y(t)$ are the displacements of the pendula. The parameter $\veps$
represents the coupling of the two pendula. The classical equations of motion
for $x(t)$ and $y(t)$ are
\begin{eqnarray}
x''(t)&=&-x(t)-\veps y(t),\nonumber\\ y''(t)&=&-y(t)-\veps x(t).
\label{e11}
\end{eqnarray}
If we solve these equations numerically, we observe Rabi oscillations (see
Fig.~\ref{F2}).

A simple, but as we will see not a good, way to create a $\cP\cT$-symmetric
system is to introduce a damping (loss) term in the $x$ equation and an
undamping (gain) term in the $y$ equation:
\begin{eqnarray}
x''(t)+ax'(t)+x(t)+\veps y(t)&=&0,\nonumber\\ y''(t)-ay'(t)+y(t)+\veps x(t)&=&0,
\label{e12}
\end{eqnarray}
where the damping/undamping parameter $a$ is positive. This system is $\cP\cT$
symmetric because $x$ and $y$ are interchanged under $\cP$ and the signs of the
damping and undamping terms reverse under time reversal $\cT$.

To solve (\ref{e12}) analytically we rewrite these equations as a system of four
first-order constant-coefficient differential equations:
\begin{eqnarray}
x'(t) &=& p(t)\nonumber,\\
p'(t) &=& -x(t)-\veps y(t)-ap(t),\nonumber\\
y'(t) &=& q(t),\nonumber\\
q'(t) &=& -y(t)-\veps x(t)+aq(t).
\label{e13}
\end{eqnarray}
We then express these differential equations compactly in matrix form:
\begin{equation}
{\bf V}'(t)={\bf M}{\bf V}(t),
\label{e14}
\end{equation}
where the matrix ${\bf M}$ and the vector ${\bf V}$ are
\begin{equation}
M=\left[\begin{array}{cccc} 0 & 1 & 0 & 0 \\ -1&-a &-\veps  & 0 \\ 0&0&0&1 \\
-\veps&0&-1&a \\ \end{array}\right]~~{\rm and}~~{\bf V}=\left[\begin{array}{c}
x(t)\\ p(t)\\ y(t)\\ q(t)\end{array}\right].
\label{e15}
\end{equation}

To solve (\ref{e15}) we must find the eigenvalues $E$ of the matrix $M$. To do
so we calculate the determinant $D$ of the matrix ${\bf M}-E{\bf I}$:
\begin{equation}
D={\rm det}\left[\begin{array}{cccc} -E &1& 0 & 0\\ -1 &-a-E &-\veps & 0\\
0&0&-E & 1\\ -\veps &0&-1 &a-E \end{array}\right].
\label{e16}
\end{equation}
The determinant $D$ is a simple quadratic form in $E^2$,
\begin{equation}
D=E^4+(2-a^2)E^2+1-\veps^2.
\label{e17}
\end{equation}
The solution to the equation $D=0$ is
\begin{equation}
E^2=\half\left(a^2-2\pm\sqrt{a^4-4a^2+4\veps^2}\right).
\label{e18}
\end{equation}

The system (\ref{e13}) exhibits oscillatory behavior if $E^2<0$. Two conditions
must be met for $E^2$ to be negative:
\begin{eqnarray}
(1)&\quad& a^4-4a^2+4\veps^2 > 0,\nonumber\\
(2)&\quad& a^2-2+\sqrt{a^4-4a^2+4\veps^2}<0.
\label{e19}
\end{eqnarray}
Condition (2) immediately gives $\veps<1$. Then, Condition (1) implies that 
\begin{equation}
a<a_{\rm crit}=2\left(1-\sqrt{1-\veps^2}\right).
\label{e20}
\end{equation}
Therefore, if the damping/undamping parameter $a$ is below this critical value,
the eigenvalues are imaginary and the system is in the unbroken-$\cP
\cT$-symmetric region.

\begin{widetext}

\begin{figure}[h!]
\begin{center}
\includegraphics[trim=0mm 10mm 0mm 5mm,clip=true,scale=0.315]{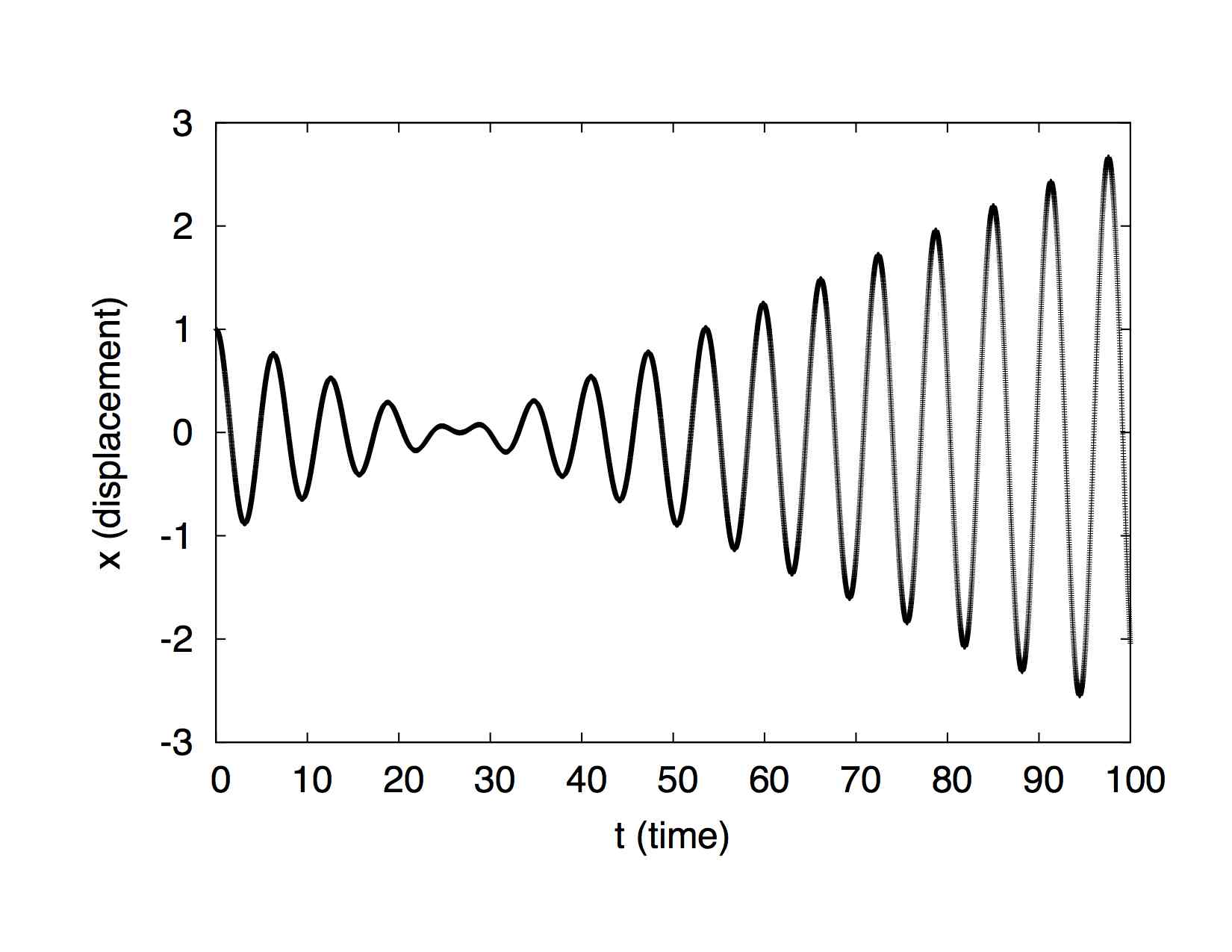}
\hspace{0.1cm}
\includegraphics[trim=0mm 10mm 0mm 5mm,clip=true,scale=0.315]{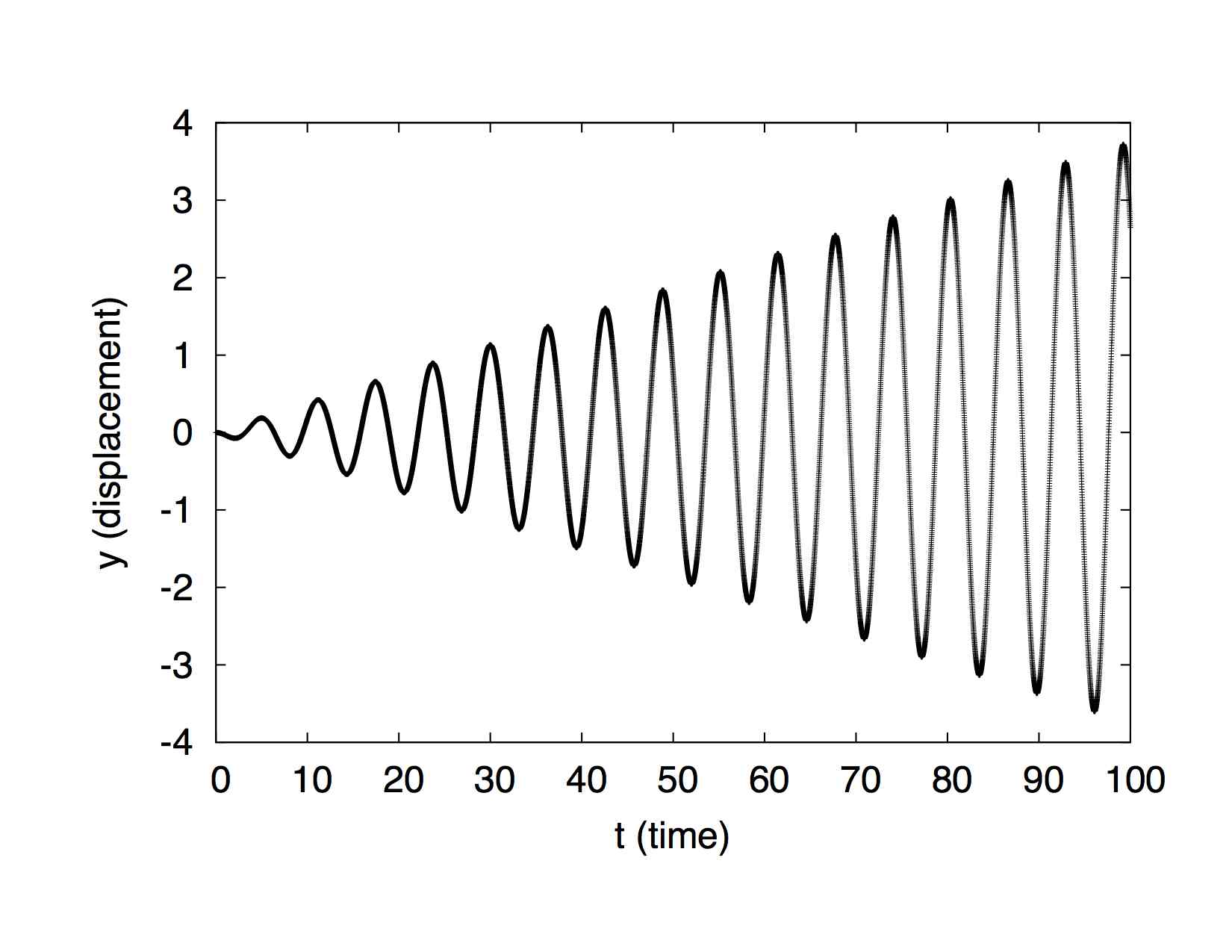}
\end{center}
\caption{Numerical solution to the overly simple coupled-oscillator equations
(\ref{e12}) for the parameter choice $a=\veps=0.075$, which is deep in the
broken-$\cP\cT$-symmetric region. The $x$ displacement is shown in the left
panel and the $y$ displacement is shown in the right panel. As expected, the $y$
oscillator exhibits gain, but after an initial period of decay, the $x$
oscillator {\it also} exhibits gain. This is because the system does not
conserve energy, and the excess energy in the $y$ oscillator leaks into the 
$x$ oscillator.}
\label{F3}
\end{figure}

\end{widetext}

As the damping/undamping parameter $a$ increases from 0, the frequency of the
Rabi power oscillations increases. Eventually, at the critical critical value of
$a$ the Rabi frequency is the same as the oscillation frequency. Above this
value of $a$ we enter the region of broken $\cP\cT$ symmetry; the Rabi
oscillations cease and we observe the onset of exponential behavior. However, a
numerical plot reveals the shortcomings of the overly simple model (\ref{e12}).
In Fig.~\ref{F3} we plot the $x(t)$ and $y(t)$ for the case $a=\veps=0.075$.
Observe that $y(t)$ grows exponentially, but after an initial period of
exponential decay $x(t)$ {\it also} grows exponentially. This is because a
balanced damping and undamping violates energy conservation. For the model
equations (\ref{e12}), the rate of growth of $y(t)$ due to undamping is
proportional to the velocity of $y(t)$ and the rate of decay of $x(t)$ is
proportional to the velocity of $x(t)$. Thus, as $y(t)$ gets big and $x(t)$
gets small, the loss of energy in the $x$ oscillator becomes insignificant while
the growth of energy in the $y$ oscillator becomes immense. Then, because the
two oscillators are coupled, $x(t)$ eventually becomes large because the energy
in the $y$ oscillator leaks into the $x$ oscillator.

\subsection{An improved energy-conserving oscillator model}
\label{ss3b}

Evidently, we need to construct a better model in which the source antenna
radiates the same amount of energy as is absorbed by the sink antenna. We do
this by constructing a pair of difference-differential equations: We begin with
the coupled-oscillator equations in (\ref{e11}) in which there is no loss or
gain. We then impose the additional condition that whenever $x(t)$ reaches an
amplitude maximum, we remove a given fixed fraction of the energy $g$ from the
$x$ oscillator (by decreasing the amplitude appropriately and reinitializing the
motion). Then, when $y(t)$ is next at a maximum we transfer this exact amount of
energy to the $y$ oscillator at its peak by increasing its amplitude
accordingly.

Numerical solutions for this improved oscillator model are shown in
Figs.~\ref{F4} and \ref{F5}. In Fig.~\ref{F4} we show what happens if we take
$\veps=0.05$ and take $g=0.01$. Here, we remove $1\%$ of the energy of the $x$
oscillator each time it reaches its maximum amplitude and then transfer this
energy to the $y$ oscillator when it reaches its next amplitude maximum; this
transfer reduces the amplitude of the $x$ oscillator at its maximum to $\sqrt{
0.99}=99.5\%$ of its former value. The coupling $\veps=0.05$ is large enough to
keep the system in the region of unbroken $\cP\cT$ symmetry. The signal that the
$\cP\cT$ symmetry is not broken is that the Rabi oscillations persist and the
amplitudes of the oscillators remain constant.

In Fig.~\ref{F5} we take the coupling smaller $\veps=0.01$ and $g$ larger $g=
0.3$. [Now, at each swing only $70\%$ of the energy in the $x$ oscillator
remains; when we make this transfer we reduce the peak amplitude to $\sqrt{0.7}=
83.7\%$ of its former value. These amplitude reductions are clearly visible in
the graph of $x(t)$.] In this case the energy transfer overwhelms the small
coupling of the oscillators and we are in the broken-$\cP \cT$-symmetric region.
The signal that the $\cP\cT$ symmetry is broken is that the Rabi oscillations
cease and the amplitudes of the $x$ (and $y$) oscillators decrease (and
increase) towards their limiting values.

\subsection{Description of the experimental setup}
\label{ss3c}

The experiment approximates the energy-conserving mathematical oscillator model
very closely. Two identical pendula are suspended from a horizontal rope of
total length 94 cm (see Fig.~\ref{F6}). By adjusting the tension in the
horizontal rope, we can vary the coupling of the two pendula. Each pendulum
consists of a 50-gram cylindrical mass (diameter $2.3$ cm, height 4.4 cm)
hanging from a string of length 39 cm. The separation of the two pendula is 34
cm.

In order to add or subtract energy from a pendulum, an electromagnet is situated
about 5 cm from top of each string. When the electromagnet is turned on, it
applies a horizontal force to a small iron nail that is attached to the string
supporting the bob. The electromagnets are only on for about 10 msec, so that
they provide a brief impulse to the string. The impulse is applied to the left
pendulum when the pendulum is moving {\it away} from the electromagnet, so the
effect of the impulse is to subtract a small amount of kinetic energy from the
pendulum. However, the impulse is applied to the right pendulum when the
pendulum is moving {\it towards} the electromagnet, so the effect of the impulse
is to add a small amount of kinetic energy to the pendulum.

Each of the electromagnets is triggered by a pair of optical sensors located
just above the pendulum bobs. For the electromagnet to fire, the two optical 
sensors must be triggered in a prescribed order, so the output of the optical
sensors is fed into a simple logic circuit.

To record the motion of the pendula we use a small video camera mounted on a
goose-neck stand \cite{r18}. The camera looks up from underneath each pendulum
and records the instantaneous position of the pendulum 15 times per second. We
then draw a curve through these data points to display the motion of the
pendulum.

We begin our experiment by taking data with the magnets turned off (see
Fig.~\ref{F7}). The tension in the horizontal rope is 200 grams. The key
characteristic feature of the swinging pendula is that they exhibit Rabi power
oscillations. The experimental data are in qualitative agreement with the
theoretical predictions in Fig.~\ref{F2}. [Note that the simple theoretical
model (\ref{e12}) with vanishing damping/undamping parameter $a$ is the same as
the improved energy-conserving oscillator model with vanishing energy-transfer
parameter $g$.] 

\begin{widetext}

\begin{figure}[t!]
\begin{center}
\includegraphics[trim=0mm 10mm 0mm 5mm,clip=true,scale=0.315]{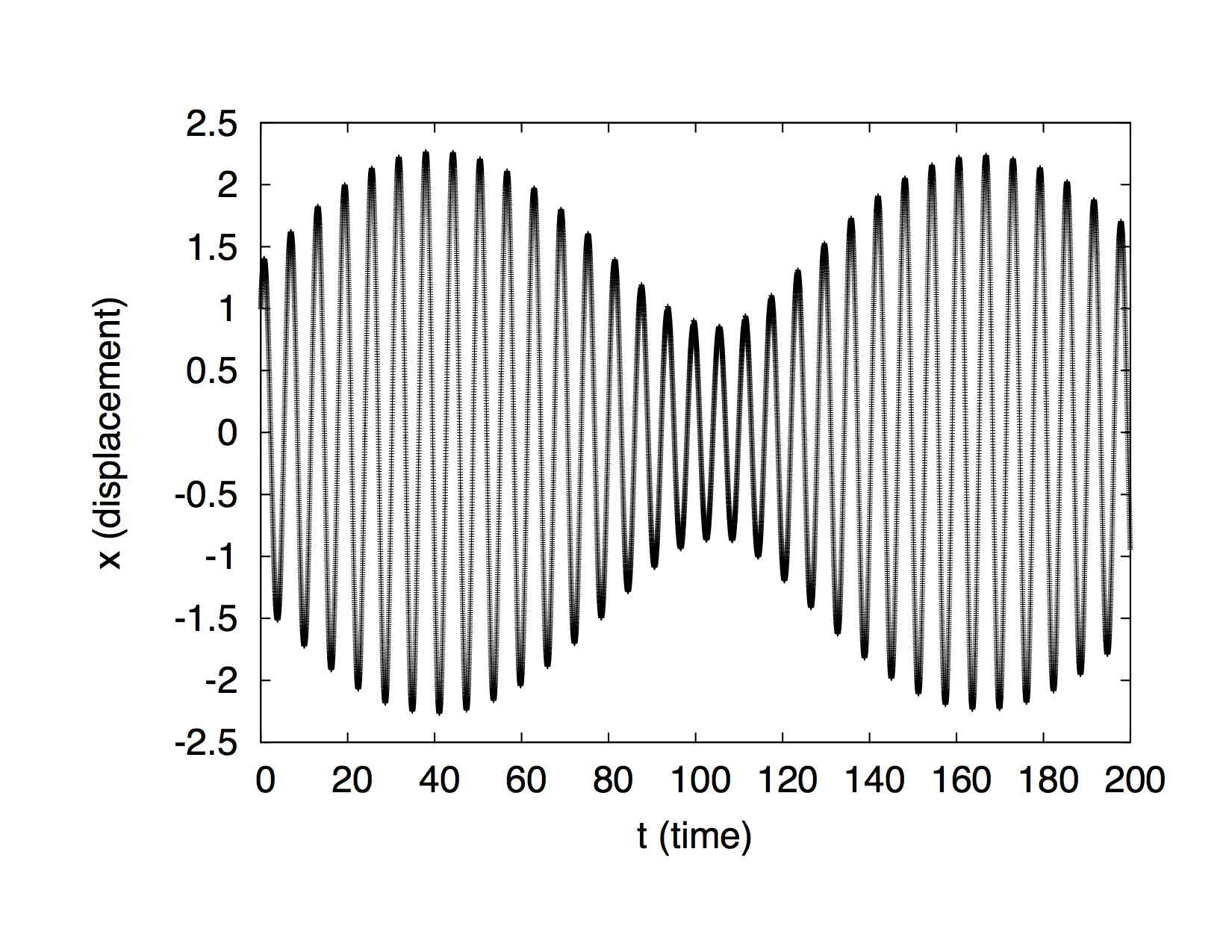}
\hspace{0.1cm}
\includegraphics[trim=0mm 10mm 0mm 5mm,clip=true,scale=0.315]{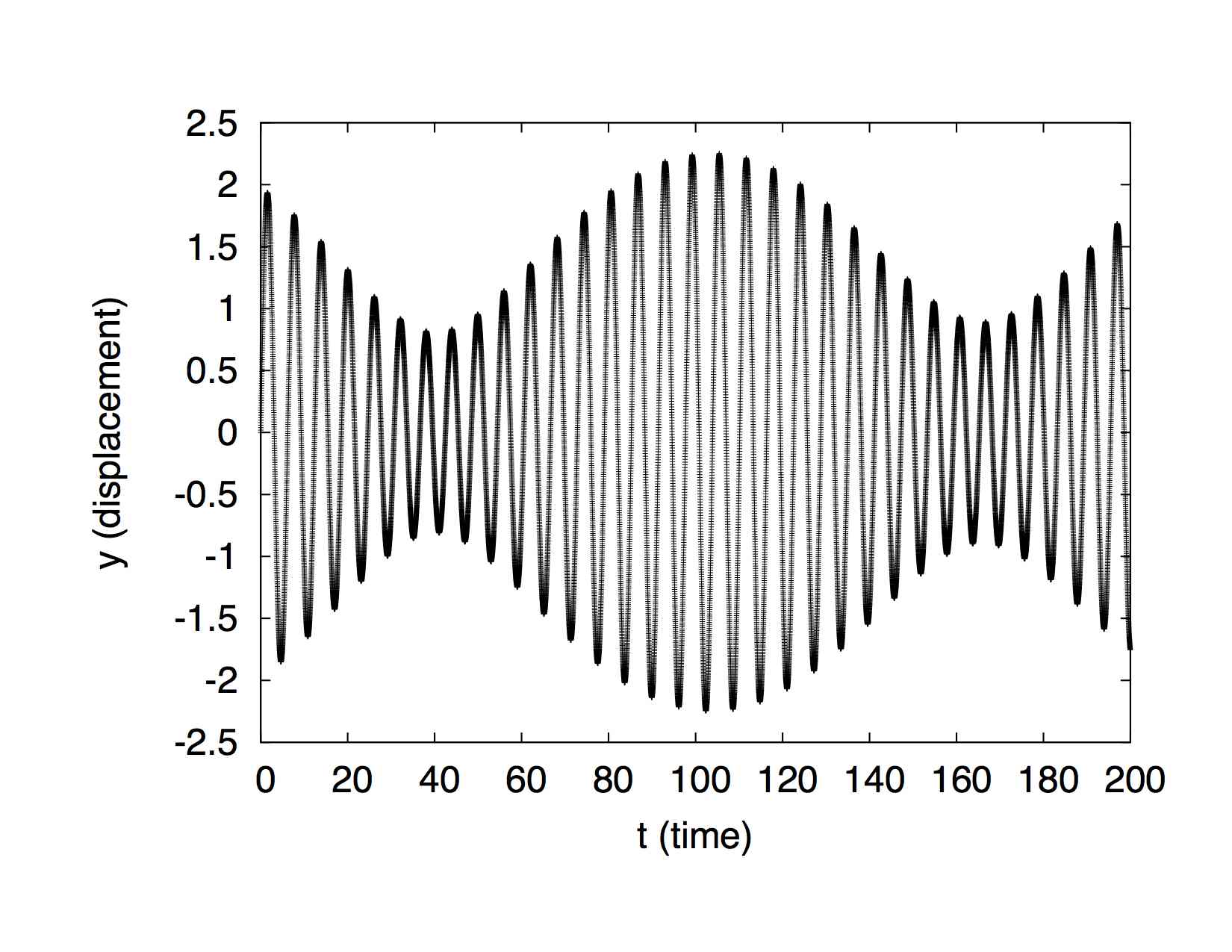}
\end{center}
\caption{Mathematical simulation of the improved energy-conserving oscillator
model in the unbroken-$\cP\cT$ region. In this graph $\veps=0.05$ and $g=0.01$.
(That is, $1\%$ of the energy in the $x$ oscillator is removed each time $x$
reaches a peak, and this exact amount of energy is then transferred to the $y$
oscillator when it reaches a peak. This transfer reduces the peak-$x$ amplitude
to $99.5\%$ of its former value, which is too small to be seen on this graph.)
The motion of the $x$ oscillator is shown in the left panel and the motion of
the $y$ oscillator is shown in the right panel. We can see that the $\cP\cT$
symmetry is not broken: The Rabi oscillations persist and the amplitudes of the
oscillators remain constant.}
\label{F4}
\end{figure}

\begin{figure}[h!]
\begin{center}
\includegraphics[trim=0mm 10mm 0mm 5mm,clip=true,scale=0.315]{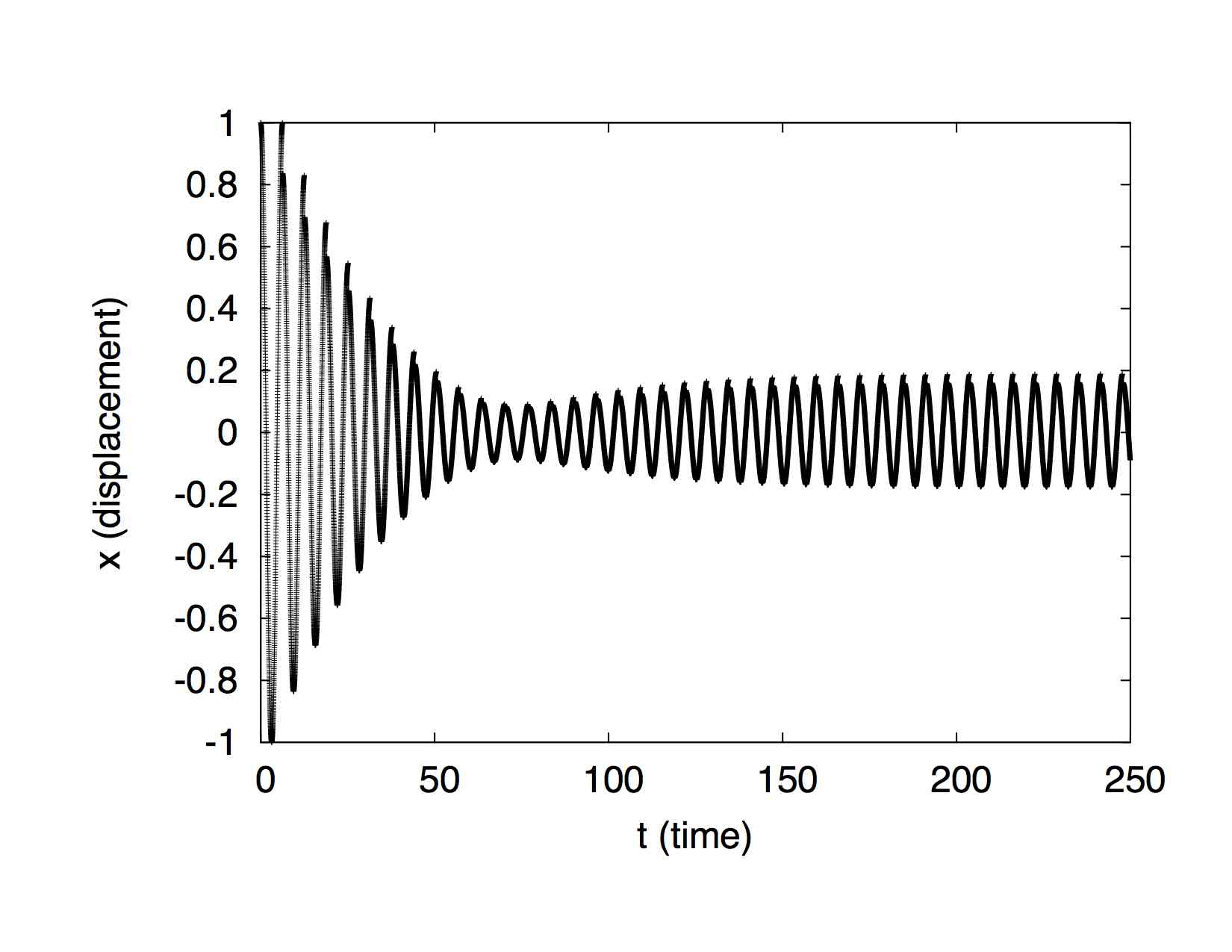}
\hspace{0.1cm}
\includegraphics[trim=0mm 10mm 0mm 5mm,clip=true,scale=0.315]{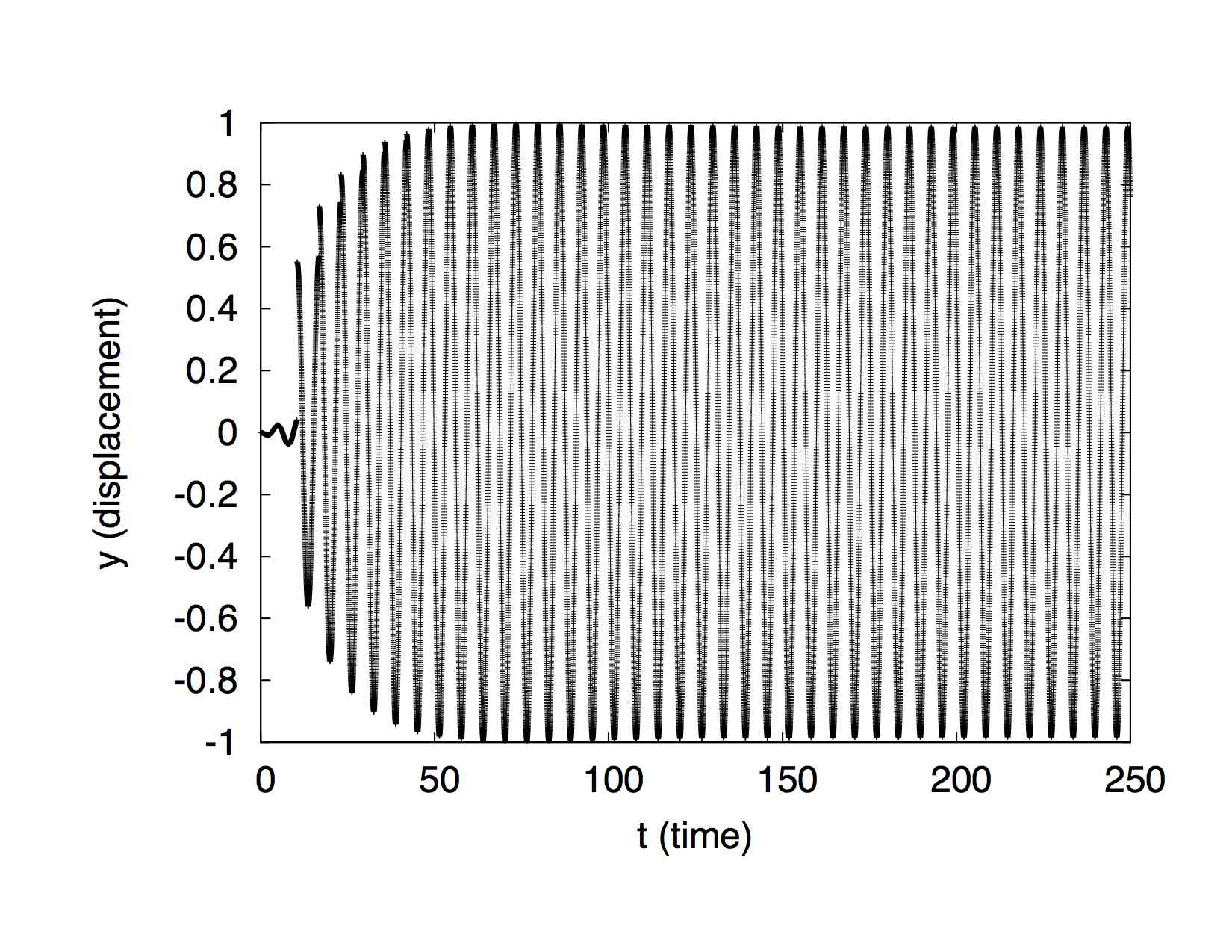}
\end{center}
\caption{Mathematical simulation of the improved energy-conserving oscillator
model in the broken-$\cP\cT$ region. The coupling here is $\veps=0.01$, which is
smaller than that in Fig.~\ref{F4}. Also, $g=0.3$, which is larger than that
in Fig.~\ref{F4}. For this value of $g$, $30\%$ of the energy in the $x$
oscillator is removed each time $x$ reaches a peak, and this exact amount of
energy is then transferred to the $y$ oscillator when it reaches a peak. This
transfer reduces the peak amplitude to $0.837$ of previous value, and this
change can be seen in the plot of $x(t)$. The motion of the $x$ oscillator is
shown in the left panel and the motion of the $y$ oscillator is shown in the
right panel. Observe that the Rabi oscillations cease and that the $x$
oscillations die down to a limiting amplitude and correspondingly the $y$
oscillations increase to a limiting amplitude. This is the characteristic
behavior of an oscillator system having a broken $\cP\cT$ symmetry.}
\label{F5}
\end{figure}

\begin{figure}[h!]
\begin{center}
\includegraphics[scale=0.85]{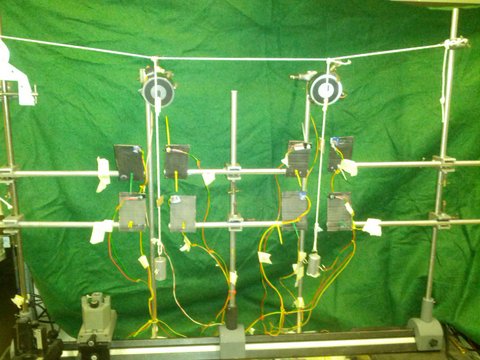}
\caption{View of the two-pendulum experiment. Two pendula are suspended from a
horizontal rope and the tension in the rope is adjusted to increase or decrease
the coupling of the pendula. The horizontal rope runs around a wheel to the left
(not shown) and is attached to a tray upon which weights can be added or
subtracted to change the tension. Electromagnets near the top of the strings
supporting the bobs apply brief impulses to small iron nails attached by white
tape to the strings. The electromagnets are triggered by pairs of optical
sensors just above the pendulum bobs. The electromagnets are timed so that on
each swing a small amount of kinetic energy is subtracted from the left pendulum
and a roughly equal amount of kinetic energy is added to the right pendulum.} 
\label{F6}
\end{center}
\end{figure}

\begin{figure}[h!]
\begin{center}
\includegraphics[scale=0.31]{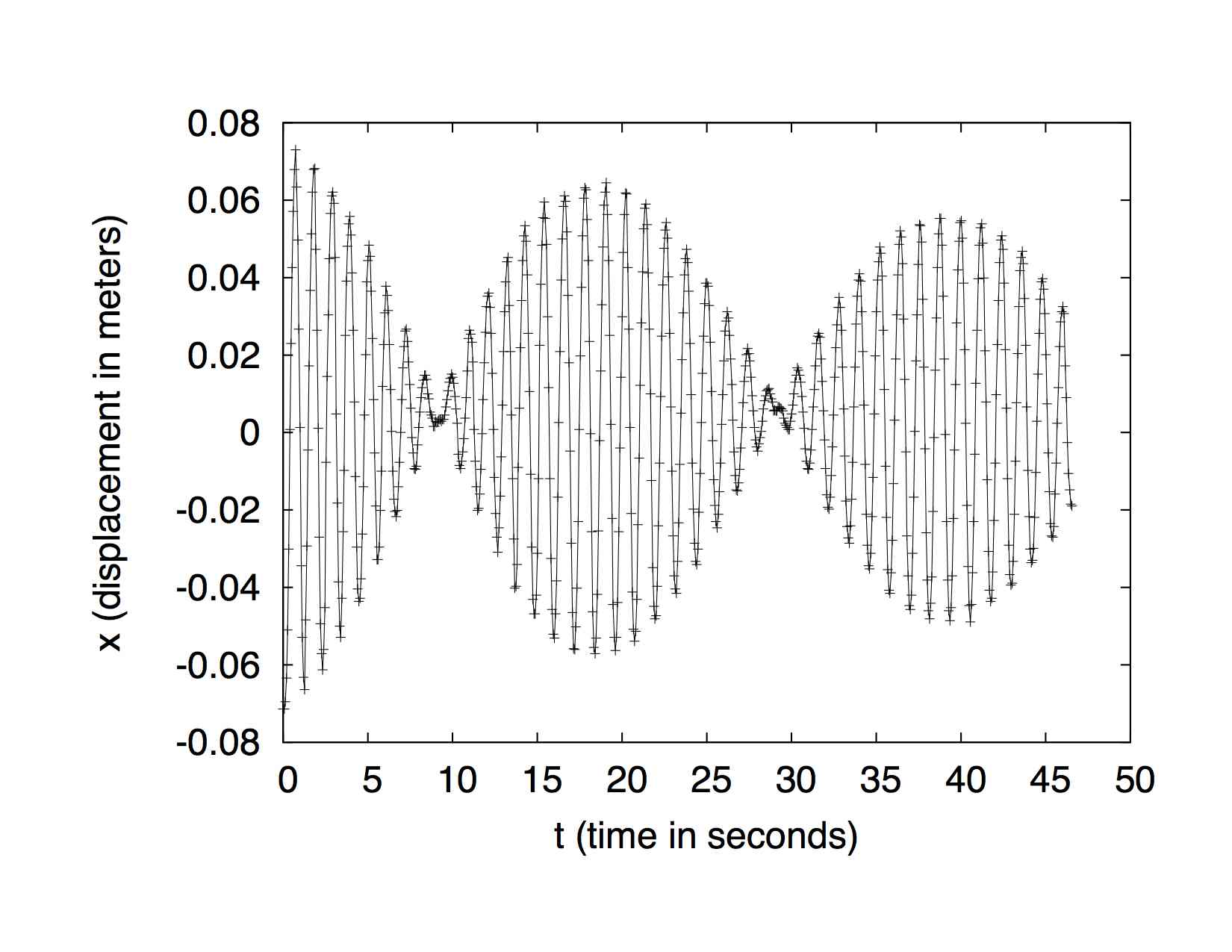}
\hspace{0.1cm}
\includegraphics[scale=0.31]{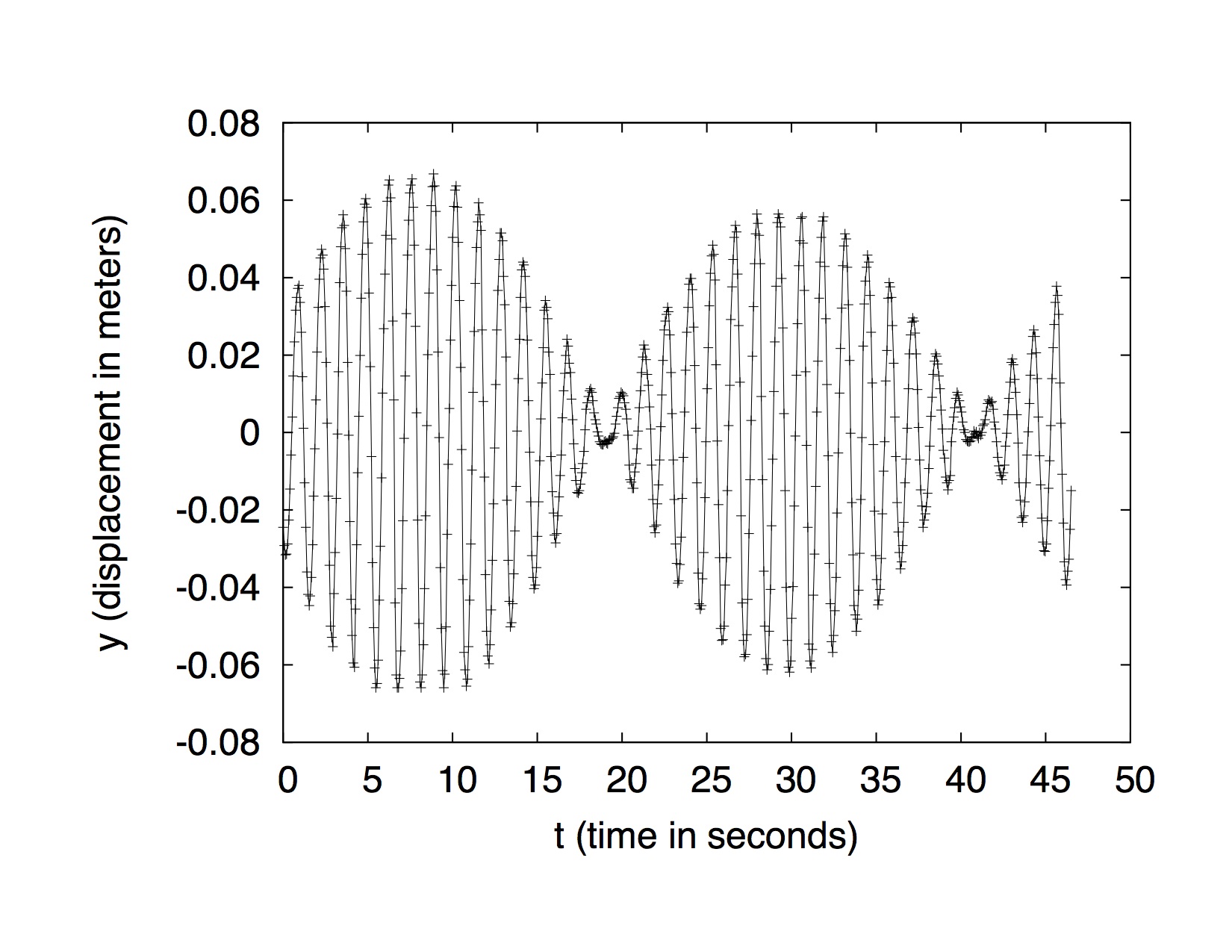}
\end{center}
\caption{Experimentally measured motion of the pendula with the magnets turned
off. The tension in the string is 200 grams. To produce these graphs we use a
camera that records the instantaneous position of each pendulum (tick marks on
the graph) 15 times per second. We then fit a curve through these data points.
The motion of the pendula is qualitatively similar to that in the theoretical
curves in Fig.~\ref{F2}. In this configuration the $\cP\cT$ symmetry is
unbroken. The signal for an unbroken $\cP\cT$ symmetry is the presence of
Rabi power oscillations, which are $90^\circ$ out of phase. One can observe a
slight decay in the amplitudes of the pendula due to friction.}
\label{F7}
\end{figure}

\end{widetext}

Next, we turn on the magnets weakly and increase the coupling of the pendula by
increasing the tension in the string from which the pendula hang from 200 to
400 grams. The experimental data are shown in Fig.~\ref{F8}. Note that the Rabi
power oscillations persist. Again, the experimental data are in qualitative
agreement with the theoretical predictions in Fig.~\ref{F4}. We conclude that we
are still in the region of unbroken $\cP\cT$ symmetry.

Finally, we make the magnets stronger and increase the tension in the string
from 400 grams to 600 grams in order to weaken the coupling of the pendula. The
experimental data is shown in Fig.~\ref{F9}. If we compare this data to the
theoretical predictions in Fig.~\ref{F5}, we again observe good qualitative
agreement: The Rabi power oscillations have ceased and we can see that the
amplitudes of the pendula level off and rapidly approach their asymptotic
values. We conclude that we have entered the region of broken $\cP\cT$ symmetry.

\section{Final remarks}
\label{s4}

In summary, the phase transition between an unbroken and a broken $\cP
\cT$-symmetric phase is easy to explain at an intuitive level. It is simply a
matter of whether two systems, one with gain and the other with loss, are
coupled strongly enough to be in equilibrium. The phase transition, which
takes place at a critical value of the coupling, can be observed in a mechanical
system consisting of two coupled pendula, one with damping and the other with
undamping. In the unbroken phase we observe periodic Rabi power oscillations.
However, when the damping/undamping becomes strong relative to the strength of
the coupling, the system enters a broken-$\cP\cT$-symmetric phase. At this
transition the Rabi power oscillations cease, and the amplitudes of the pendula 
approach limiting values.

A natural extension of the experimental work described in this paper to
investigate $\cP\cT$-symmetric chains of coupled pendula with alternating
damping and undamping. It is predicted that an optical system consisting of
coupled optical fibers with alternating loss and gain will exhibit birefringence
\cite{r19}. Our objective in studying this multiple pendulum system is to
observe the mechanical equivalent of birefringence.

We have also begun to study $\cP\cT$-symmetric systems of coupled accoustic wave
guides. We have already found that coupled accoustic wave guides with
alternating loss and gain can exhibit a rich and elaborate array of phase
transitions.

CMB thanks the U.K.~Leverhulme Foundation and the U.S.~Department of Energy for
financial support.
\begin{widetext}

\begin{figure}[h!]
\begin{center}
\includegraphics[scale=0.315]{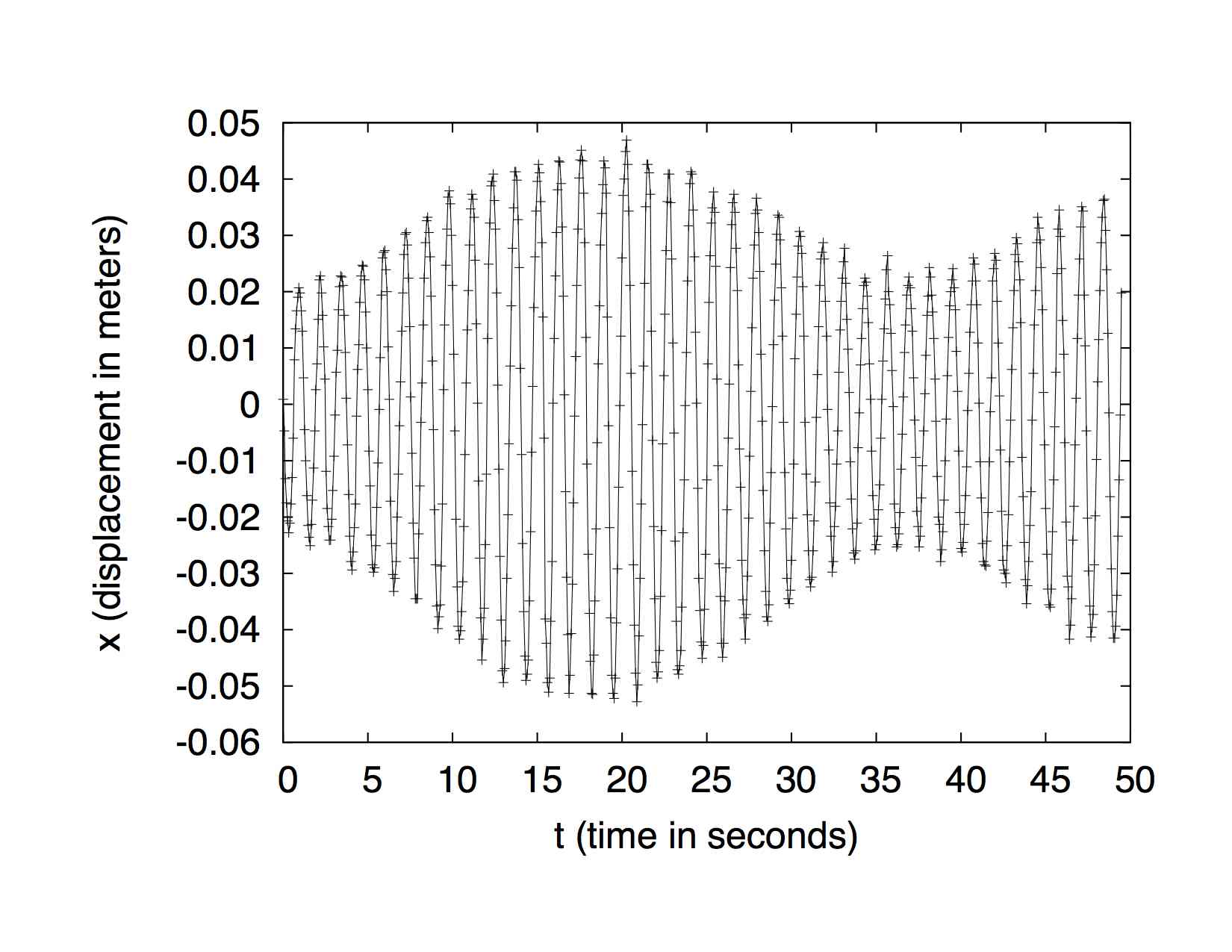}
\hspace{0.1cm}
\includegraphics[scale=0.315]{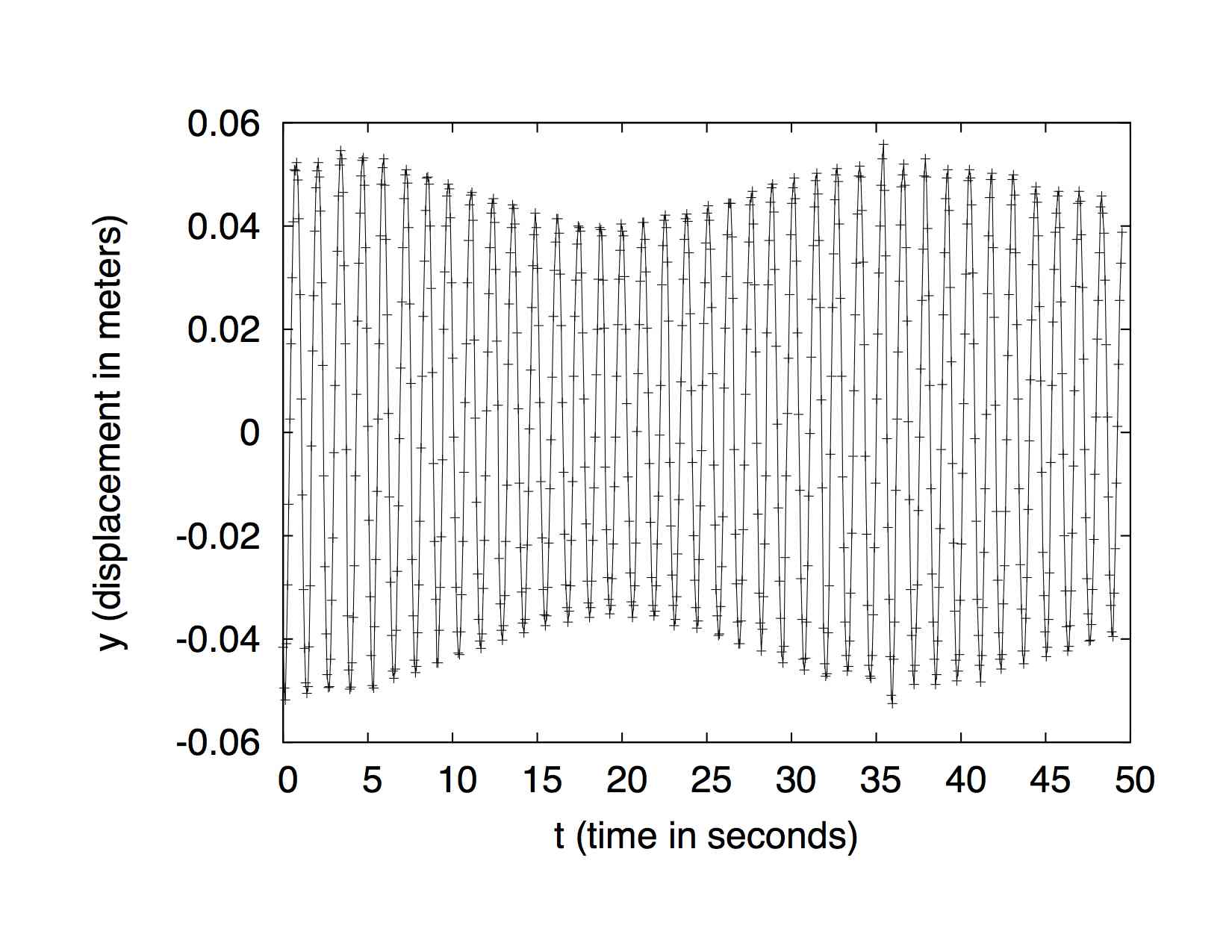}
\end{center}
\caption{Experimental data showing the motion of the pendula when the magnets
are turned on weakly, so that there is a weak loss and gain, and the pendula are
strongly coupled (the tension in the supporting rope is lowered to 400 grams).
Observe that the Rabi power oscillations in Fig.~\ref{F7} persist. This
means that the system is in a region of unbroken $\cP\cT$ symmetry.}
\label{F8}
\end{figure}

\begin{figure}[h!]
\begin{center}
\includegraphics[scale=0.315]{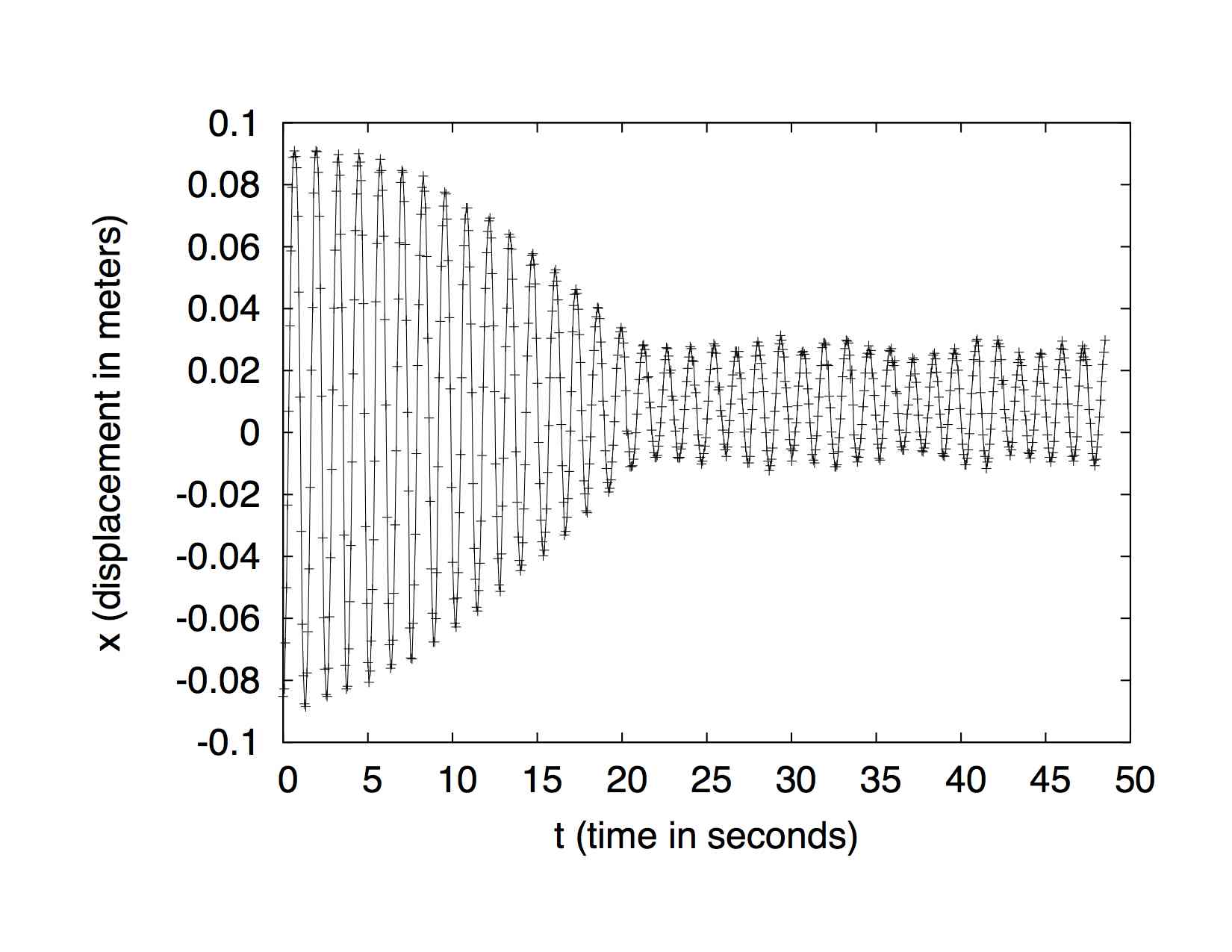}
\hspace{0.1cm}
\includegraphics[scale=0.315]{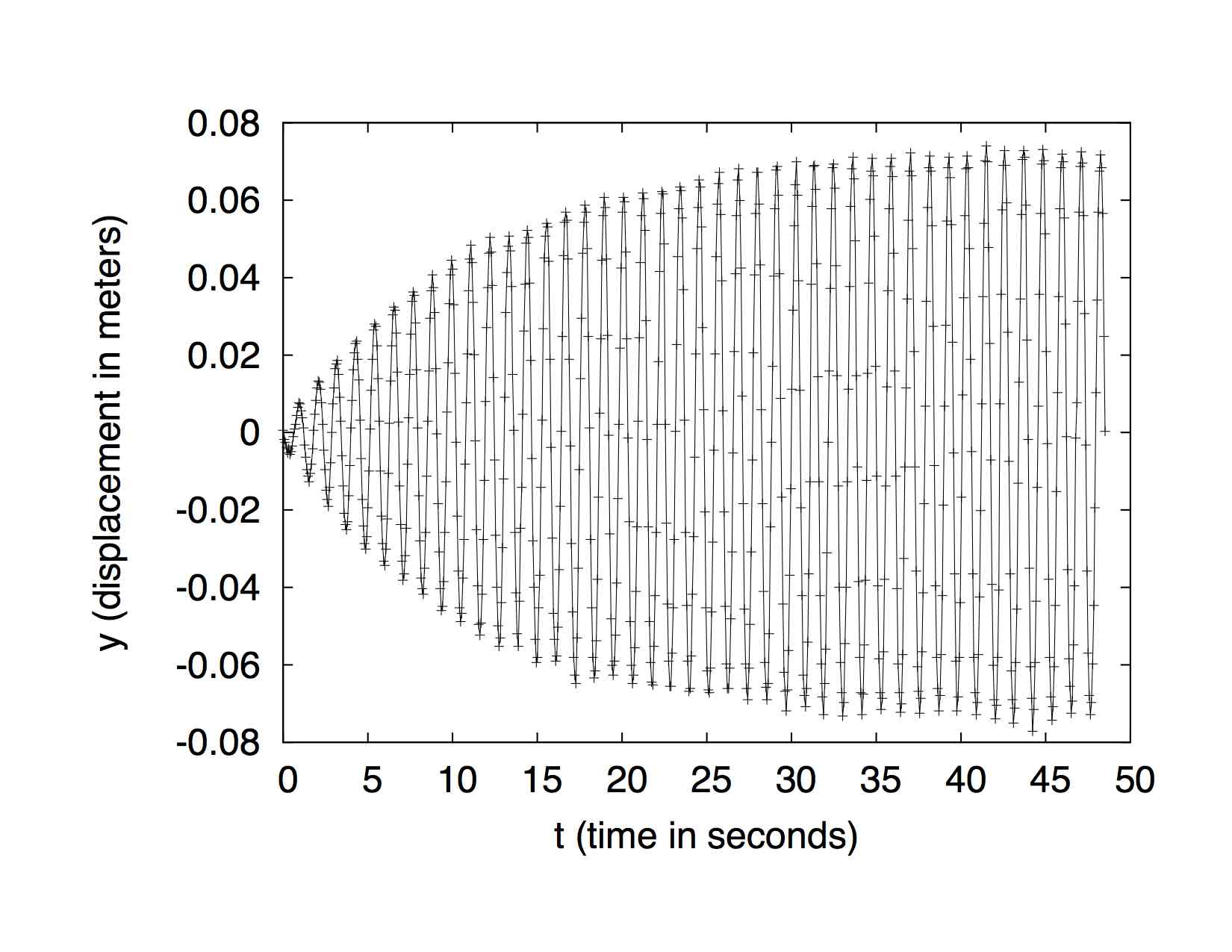}
\end{center}
\caption{Experimental data showing the motion of the pendula when the magnets
are turned on strongly and the coupling of the pendula is weak (the tension in
the supporting string is raised to 600 grams). Observe the the Rabi oscillations
have ceased. This is the signal that the system is in a region of broken $\cP
\cT$ symmetry.}
\label{F9}
\end{figure}

\end{widetext}

\end{document}